\documentclass[pre,twocolumn,showpacs,preprintnumbers,amsmath,amssymb]{revtex4}

\usepackage{graphicx} 
\usepackage{dcolumn}
\usepackage{bm}
\usepackage{psfrag}
\usepackage{epsfig}

\begin{document}

\title{Projection operator approach to lifetimes of electrons in metals}

\author{Mehmet Kadiro\=glu}
\email{mkadirog@uos.de}
\author{Jochen Gemmer}
\email{jgemmer@uos.de}
\affiliation{Department of Physics, University of Osnabr\"uck,
D-49069 Osnabr\"uck, Germany}

\date{\today}

\begin{abstract}
We present an alternative approach to the calculation of the lifetime
of a single excited electron (hole)  which interacts with the Fermi sea of
electrons in a metal. The metal is modelled on the level of a Hamilton
operator comprising a pertinent  dispersion relation and scattering term.
To determine the full relaxation dynamics we
employ an adequate implementation of the  time-convolutionless projection operator
method (TCL). This yields an analytic expression for the decay rate
which  allows for an intuitive interpretation in terms of scattering
events. It may furthermore be efficiently evaluated by means of a
Monte-Carlo integration scheme. As an example we investigate aluminium
using, just for simplicity, a jellium-type model. This way we obtain data
which are  directly comparable to results from a self-energy formalism. Our
approach applies to arbitrary temperatures.
 \end{abstract}

\pacs{72.10.-d, 72.10.Bg, 72.15.Lh}
\keywords{}
\maketitle


\section{Introduction}
\label{intro}

For several decades, the dynamics of excited electrons in metals has been the subject of intense
research in theoretical and experimental solid state
physics.
\cite{Quinn1,Quinn2,baym,Bauer1,Bauer4,Bauer2,Echenique1,Echenique2,Hofer,Jaeckle}.
These investigations are motivated by the fact that a lot
of physical and chemical properties of
metallic materials depend essentially on those dynamics \cite{Pines,Mahan,Kittel}. Lifetimes of (photo)excited electrons in metals
are always short (on the order of femtoseconds) but the immense
progress in ultrafast laser technology now allows for an experimental
determination of such times, corresponding investigations are ongoing \cite{Bauer3}.\\    

Today a number of methods are used to calculate lifetimes of electrons. 
Practically all of them are formulated within the framework of Green's functions (many-body theory) 
and aim at determining the self-energy, particularly its imaginary part \cite{Fetter,Kirzhnits,Brown}. 
Many of them employ a screened interaction (``W'') and a truncated expansion of the 
self energy in terms of this screened interaction (``GW-approximation'') 
\cite{Hedin1,Hedin2,Echenique1,Echenique2,Chulkov1,Chulkov2,Rubio,Schoene,Rolli1,Rolli2,Mari,Nechaev}. The 
screened interaction is frequently 
obtained through a ``random phase approximation'' (RPA) \cite{Pines}. For a simple, sufficiently dense, homogeneous gas of electrons interacting 
through coulomb repulsion (jellium model) an approach along the above scheme is even 
feasible analytically and yields a closed expression for the lifetimes close to the Fermi edge (see below) \cite{Quinn1,Quinn2}. 
In a certain sense (which is described in more detail below (\ref{HamJel})) 
this approach leads to lifetimes which may quantitatively be compared to experimental data on e.g., aluminium \cite{Bauer1,Schoene,Nechaev}.
Of course timely state of the art approaches go beyond jellium and 
exploit not only the  traditional self-energy formalism 
but also density functional methods, etc. \cite{Schoene,Rolli1,Rolli2,Mari}.\\
Our approach is, in contrary, not based on Green's functions at 
all but on projection operator techniques. A main motivation of our work is to 
demonstrate that a pertinent projective approach 
\cite{Nakajima,Zwanzig,Grabert,Fulde,Rau,Burnett,Neu,Breuer} is also 
capable 
to produce quantitative results on lifetimes. 
Our central formula from which the lifetimes are eventually calculated is in accord with expressions 
that may be derived within the above many-body approach (see below (\ref{final})). Furthermore  it allows for an interpretation 
in terms of scattering events. This encourages a further development of projection techniques as alternative quantitative 
tools for the investigation of relaxation and transport dynamics in condensed matter systems (find more on this at the end of Sec. \ref{sec-3}). 
However, 
our approach starts form an effective model comprising  pertinent quasi-particle dispersion relations and an appropriate 
screened interaction. The (generically subtle) provision of such a suitable effective model 
is not part of our present analysis, the effective model thus has to be supplied by other means.\\
The article at hand is organized as follows: In Sec. \ref{sec-1} we
give a very brief introduction to the
time-convolutionless projection operator method \cite{Breuer} and
apply it to a general interacting quantum gas thus obtaining an expression for the electronic lifetime. 
In Sec. \ref{sec-2} we evaluate this expression numerically 
for a ``screened'' jellium model tuned to describe aluminium. We compare 
our results to other available data and comment on computing times.
Eventually we close with discussion, summary and outlook.

\section{ projective  approach to  occupation number dynamics in interacting
  quantum gases}
\label{sec-1}
\begin{figure}
\centering
\includegraphics[width=80mm]{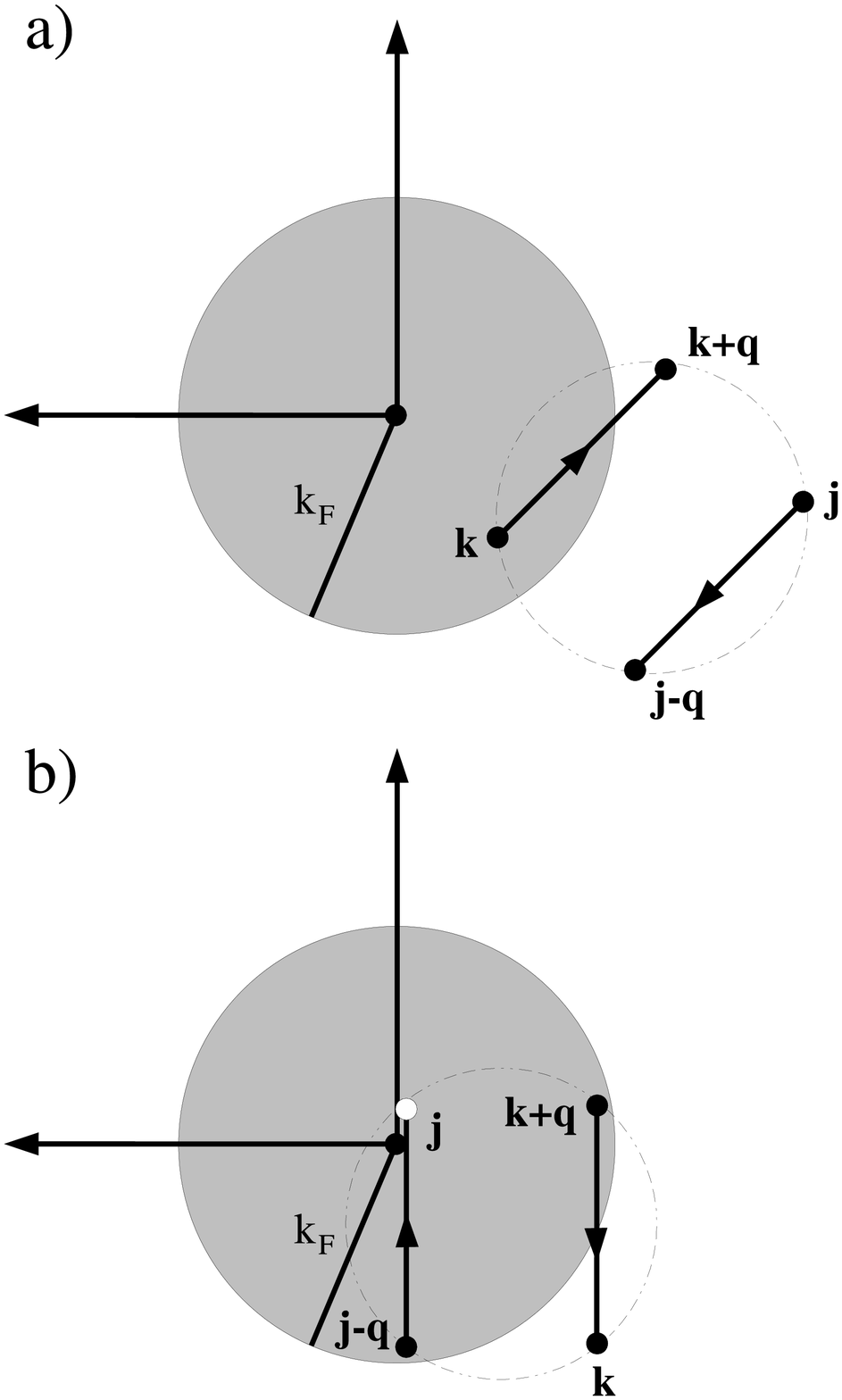}
\vspace*{1mm}
\caption{\label{Collision}Schematic representation of the  underlying collision processes in momentum space  as described in the text ($k_{F}$ denotes the Fermi 
momentum).\\
 a) A collision process through which an excited electron at momentum ${\bf j}$ vanishes from its initial momentum mode.\\
b) A collision process "filling" a hole within the Fermi sphere at  ${\bf j}$.\\
The dashed circles denote the possible outgoing momenta under momentum and energy conservation.}
\end{figure}

To determine the lifetime of an electron initially occupying some
momentum eigenstate we analyze the dynamics of the corresponding
occupation number. A formalism which allows for such an analysis is
the TCL-method \cite{Breuer}. In general the latter is a perturbative  projection
operator technique which produces autonomous equations of motion for
the variables of interest (``relevant information''). The technique
may be applied to quantum system with a Hamiltonian of the
type $\hat{H}=\hat{H}_{0}+\lambda\hat{V}$ where
$\lambda$ has to be in some sense small \cite{Breuer}
In order to apply this method one first has to construct a suitable
projection operator $\mathcal{P}$.
Formally, this is a linear map which projects any density matrix $\varrho (t)$ to a matrix
$\mathcal{P}\varrho (t)$ that is determined by a certain set of
variables. These variables should match with the variables of interest. Moreover $\mathcal{P}$ has to fulfill the property
of a projection operator, that is $\mathcal{P}^{2}=\mathcal{P}$.
For initial states with $\mathcal{P}\varrho (0)=\varrho (0)$ the TCL scheme leads to a time-local differential equation for the dynamics of 
$\mathcal{P}\varrho$:
\begin{eqnarray}
\partial_{t}\mathcal{P}\varrho(t)=\Gamma (t)\mathcal{P}\varrho(t), \hspace{0.5cm} \Gamma
(t)=\sum\limits_{k=1}^{\infty}\lambda^k\Gamma_{k}(t),
\label{expva}
\end{eqnarray}
where the perturbative expansion used in the last equations is in
principle exact.
However, for a description to leading order,  which is typically and in our
case the second order, one has to determine $\Gamma_{2} (t)$. Whether
or not a leading order description will yield a reasonable result is a
somewhat subtle question \cite{Christian} but the expansion is well controlled and systematic, i.e, in principle higher order terms could be incorporated in a straightforward manner \cite{Breuer}. A widely accepted
indicator  for the validity of the truncation is a clear timescale
separation between the resulting relaxation dynamics and the decay of
the correlation function, the latter being introduced below. However,
here we are going to  focus on the leading order and comment on the
timescales below when we eventually arrive at concrete lifetimes.
In the
literature \cite{Breuer} one finds 
\begin{eqnarray}
\Gamma_{2} (t)=\int\limits_{0}^{t}dt'\mathcal{P}\mathcal{L}(t)\mathcal{L}(t')\mathcal{P},
\label{Cum2}
\end{eqnarray}
with $\mathcal{L}(t)=\frac{\imath}{\hbar}[\hat{V}(t),\#]$, where $\#$
denotes a placeholder for an operator which shall be inserted into the
commutator. $V(t)$ refers to a perturbation in the interaction picture.
With (\ref{expva}) and (\ref{Cum2}) we obtain:
\begin{eqnarray}
\partial_{t}\mathcal{P}\varrho(t)=\int\limits_{0}^{t}dt'\mathcal{P}\mathcal{L}(t)\mathcal{L}(t')\mathcal{P}\varrho(t).
\label{equ1}
\end{eqnarray}
Now for an  concrete application we have to specify the underlying
quantum  model and a suitable projection operator.\\
The systems we investigate are interacting quantum gases, here of the
``spinless fermions''- type. The corresponding Hamiltonians may be written as 
\begin{eqnarray}
\underbrace{\sum\limits_{{\bf k}}\varepsilon_{{\bf k}}a_{{\bf k}}^{\dagger}a_{{\bf 
k}}}_{\hat{H}_{0}}+\underbrace{\frac{1}{2}\sum\limits_{{\bf 
k},{\bf l},{\bf q}}V({\bf q})a_{{\bf
k+q}}^{\dagger}a_{{\bf
l-q}}^{\dagger}a_{{\bf l}}a_{{\bf k}}}_{\hat{V}},
\label{Hamilton1}
\end{eqnarray}
where $\varepsilon_{{\bf k}}$ denotes a dispersion relation of free particles and $V({\bf q})$ the matrix elements of an interaction which depends on 
the concrete system. This Hamiltonian is of the above mentioned form as
long as the interaction term $\hat{V}$ is in adequate sense
``small'' (see below). As will be demonstrated below (cf. Sec. \ref{sec-2}) it is thus reasonable 
to choose for $\varepsilon_{{\bf k}}$ pertinent quasi-particle dispersion relations 
and particularly for  $V({\bf q})$ an adequate screened interaction. 
Note that we neglect the spin quantum
number since the system we consider in this work is paramagnetic and without 
any magnetic fields the dispersion relation is the same for both spin channels. 
Below we are going to take care of this ``spin degeneracy'' in a very simple 
form (cf. text following (\ref{lebensdauer1}). For the non-interacting 
many-particle system we may directly write down the
wavenumber (momentum) dependent ``single particle equilibrium density operator'' as:
\begin{eqnarray}
\varrho_{{\bf j}}^{eq}:=f_{{\bf j}}(\mu,T)a_{\bf j}^{\dagger}a_{\bf j}+(1-f_{{\bf j}}(\mu,T))a_{\bf j}a_{\bf j}^{\dagger},
\end{eqnarray}
with $f_{{\bf j}}(\mu,T)=(\exp ((\varepsilon({\bf j})-\mu)/k_{B}T)+1)^{-1}$ being the Fermi distribution. Since we are interested in temperature 
regimes close to $T=0$K but still $T\neq 0$K we can set the chemical potential $\mu$ $\approx$ $\varepsilon_{F}$. Further we abbreviate $f_{{\bf 
j}}(\varepsilon_{F},T)$ as $f_{{\bf j}}$.
The equilibrium density operator, again for the non-interacting case,
of the total system, $\varrho^{eq}$, may be written as the tensor product of the single particle 
density operators, i.e.
\begin{eqnarray}
\varrho^{eq}:=\bigotimes\limits_{{\bf i}}\varrho_{{\bf
    i}}^{eq},\qquad\text{note also}\qquad \tilde{\varrho}:=\bigotimes\limits_{{\bf 
i}\neq {\bf j}}\varrho_{{\bf i}}^{eq}.
\label{density}
\end{eqnarray}
Here, for later reference, $\tilde{\varrho}$ denotes the total density operator of 
the system which does not contain the subspace with respect to the
momentum mode ${\bf j}$, i.e. it is
$\varrho^{eq}=\tilde{\varrho}\otimes\varrho_{{\bf j}}^{eq}$. We
should, also for later reference,  mention here that while $\varrho^{eq}$ is strictly speaking
just the equilibrium state of the non-interacting system, it is
routinely considered to describe the single particle properties of the
weakly interacting system more or less correctly. Thus, if single
particle observables relax towards equilibrium due to the
interactions (scattering), we expect them to relax towards values
corresponding to $\varrho^{eq}$.\\
For the investigations of the dynamics of excited states we define an
operator $\Delta_{{\bf j}}$ (for the remainder of this paper $|{\bf j}\rangle$ denotes 
the excited state) as
\begin{eqnarray}
\Delta_{{\bf j}}:=(1-f_{{\bf j}})a_{\bf j}^{\dagger}a_{\bf j}-f_{{\bf j}}a_{\bf j}a_{\bf j}^{\dagger}=
a_{\bf j}^{\dagger}a_{\bf j}-f_{{\bf j}},
\end{eqnarray}
which describes the deviation of the mode occupation number $n_{\bf j}=a_{\bf j}^{\dagger}a_{\bf j}$ from its thermal equilibrium.
Now, in order to apply the TCL method to this model we construct a suitable projector as follows:
\begin{eqnarray}
\mathcal{P}\varrho(t)=\varrho^{eq}+\frac{1}{\sigma^{2}_{{\bf j}}}\text{Tr}\{\Delta_{{\bf j}}\varrho(t)\}\tilde{\varrho}\otimes\Delta_{{\bf 
j}},
\label{proj}
\end{eqnarray}
with $\varrho(t)$ being the density operator which describes the actual state of the system, $d_{{\bf j}}(t):=$Tr$\{\Delta_{{\bf j}}\varrho(t)\}$
denotes the time dependent expectationvalue of $\Delta_{{\bf j}}$ and $\sigma^{2}_{{\bf j}}:=(1-f_{{\bf j}})^{2}+f_{{\bf j}}^{2}=$Tr$\{\Delta_{{\bf 
j}}^{2}\}$.  
It is straightforward to show that with the above definitions $\mathcal{P}$ is a projector and fulfills 
$\mathcal{P}^{2}\varrho(t)=\mathcal{P}\varrho(t)$. 
Note that Tr$\{\varrho^{eq}\Delta_{{\bf j}}\}=0$ and Tr$\{\Delta_{{\bf j}}\tilde{\varrho}\otimes\Delta_{{\bf j}}\}=\sigma^{2}_{{\bf j}}$.
Before we eventually concretely apply (\ref{equ1}) to our model we
make the following approximation for an expression that appears in the
computation of (\ref{equ1}):
\begin{eqnarray}
\mathcal{L}(t')\mathcal{P}\varrho(t)&=&\frac{\imath}{\hbar}[\hat{V}(t'),\varrho^{eq}]+\frac{\imath}{\hbar}
[\hat{V}(t'),\tilde{\varrho}\otimes\Delta_{{\bf j}}]\frac{d_{{\bf j}}(t)}{\sigma^{2}_{{\bf j}}}\nonumber\\
&\approx&\frac{\imath}{\hbar} [\hat{V}(t'),\tilde{\varrho}\otimes\Delta_{{\bf j}}]\frac{d_{{\bf j}}(t)}{\sigma^{2}_{{\bf j}}},
\label{approx1}
\end{eqnarray}
The neglected commutator term essentially describes the dynamics of the
equilibrium state of the non-interacting system. Eventually we are
interested in a single particle observable. As already mentioned
above, the equilibrium state of the non-interacting system is believed
to
reasonably describe single particle observables in equilibrium even
for weakly interacting systems. Since an equilibrium state is
constant, the above commutator should not significantly contribute to
the relevant dynamics, thus we drop it. Keeping the term and
performing all following steps yields eventually an expression which
can explicitly shown to be indeed negligible in the weak coupling
limit. For clarity and briefness we omit this calculation here.\\
If we apply now (\ref{equ1}) to (\ref{proj}) and make use of (\ref{approx1}) we obtain
\begin{eqnarray}
&~&\partial_{t}\mathcal{P}\varrho(t)=\frac{1}{\sigma^{2}_{{\bf j}}}\partial_{t}d_{{\bf 
j}}(t)\cdot\tilde{\varrho}\otimes\Delta_{{\bf 
j}}=\int\limits_{0}^{t}dt'\varrho^{eq}\nonumber\\
&-&\int\limits_{0}^{t}dt'\frac{1}{\sigma^{4}_{{\bf j}}\hbar^{2}}\text{Tr}\{[\Delta_{{\bf 
j}}[\hat{V}(t),[\hat{V}(t'),\tilde{\varrho}\otimes\Delta_{{\bf j}}]]\}\tilde{\varrho}\otimes\Delta_{{\bf j}} 
d_{{\bf 
j}}(t).\nonumber\\
\label{equ2}
\end{eqnarray}
Multiplying both sides of (\ref{equ2}) with $\Delta_{{\bf j}}$ and taking the trace leads to:
\begin{eqnarray}
\partial_{t}d_{{\bf j}}(t)=
-\underbrace{\frac{1}{\hbar^{2}\sigma^{2}_{{\bf j}}}\int\limits_{0}^{t}dt'\text{Tr}\{[\Delta_{{\bf 
j}}[\hat{V}(t),[\hat{V}(t'),\tilde{\varrho}\otimes\Delta_{{\bf j}}]]\}}_{\Gamma_{\bf j}(t)}d_{{\bf 
j}}(t),\nonumber\\
\label{rateequa}
\end{eqnarray}
where $\Gamma_{\bf j}(t)$ appears as a  time-dependent damping
rate of the mode ${\bf j}$. If  $\Gamma_{\bf j}(t)$ turns out
to be approximately time-independent, the usual exponential relaxation
results.
With the substitution $t'=t-\tau$ and exploiting $[\tilde{\varrho}\otimes\Delta_{{\bf j}},\hat{H}_{0}]=[\Delta_{{\bf 
j}},\hat{H}_{0}]=0$ as well as some trace properties, we obtain for the rate:
\begin{eqnarray}
\Gamma_{\bf 
j}(t)=\frac{1}{\hbar^{2}\sigma^{2}_{{\bf j}}}\int\limits_{0}^{-t}d\tau\underbrace{\text{Tr}\{[\hat{V}(\tau),\tilde{\varrho}\otimes\Delta_{{\bf 
j}}]\cdot[\hat{V}(0),\Delta_{{\bf j}}]\}}_{C(\tau)},
\label{Rate}
\end{eqnarray}
where $C(\tau)$ denotes the correlation function which is real due to
the fact that both commutators are hermitian. The concrete evaluation
of this expression with respect to our model is straightforward but
somewhat lengthy. Thus the full computation is given in the Appendix,
here we only give and discuss the results 
After exploiting the commutators within the trace we finally obtain 
for the rate:
\begin{eqnarray}
\Gamma_{\bf j}(t,T)&=&\frac{1}{\tau_{{\bf j}}}=-\frac{2}{\hbar^{2}\sigma^{2}_{{\bf j}}}\sum\limits_{{\bf k},{\bf q}}\int\limits_{0}^{-t}d\tau|V({\bf 
q})|^{2}F({\bf 
k},{\bf 
q},{\bf j},T)\nonumber\\
&\times&\cos\left((\omega_{{\bf k}+{\bf q}}+\omega_{{\bf j}-{\bf q}}-\omega_{{\bf k}}-\omega_{{\bf 
j}})\tau\right)\nonumber
\end{eqnarray}
\vspace{-0.5cm}
\begin{eqnarray}
&=&\frac{2}{\hbar^{2}\sigma^{2}_{{\bf j}}}\sum\limits_{{\bf k},{\bf q}}|V({\bf q})|^{2}F({\bf k},{\bf q},{\bf j},T)
\cdot t\cdot\text{sinc}\left(\omega({\bf k},{\bf q},{\bf j})t\right),\nonumber\\
\label{final}
\end{eqnarray}
with:
\begin{eqnarray}
\omega({\bf k},{\bf q},{\bf j})&:=&\omega_{{\bf k}+{\bf q}}+\omega_{{\bf j}-{\bf q}}-\omega_{{\bf k}}-\omega_{{\bf
j}},\nonumber\\
\nonumber\\
F({\bf k},{\bf q},{\bf j},T)&:=&
\underbrace{(1-f_{{\bf j}})(1-f_{{\bf j-q}})(1-f_{{\bf k+q}})f_{{\bf k}}}_{F_{1}({\bf k},{\bf q},{\bf j},T)}\nonumber\\
\nonumber\\
&+&\underbrace{f_{{\bf j}}f_{{\bf j-q}}f_{{\bf k+q}}(1-f_{{\bf k}})}_{F_{2}({\bf k},{\bf q},{\bf j},T)}\nonumber\\
\end{eqnarray}
and sinc$(\omega t)$ denotes the sinus cardinalis. Obviously the
integral $\int_{-\infty}^{+\infty}  t\cdot$sinc$(\omega t)d\omega$ 
is independent of $t$. Furthermore the function gets more and more
peaked with increasing $t$ such that, as wellknown,
\begin{eqnarray}
\lim_{t\rightarrow\infty}\frac{\sin (\omega t)}{\omega}=\pi\delta(\omega).
\end{eqnarray}
Hence, since dispersion relations are smooth functions of the
wavenumber, we expect the rate  $\Gamma_{\bf j}(t,T)$ to become indeed
time-independent for times larger than $\tau_c$, if $1/\tau_c$ is an
energyscale on which dispersion relations may be linearized. Thus for
times $t$ larger than $\tau_c$ we may with good precision approximate (here we neglect the factor $\sigma^{-2}_{{\bf j}}$ since
for temperatures $T\approx0$K it is $\sigma^{-2}_{{\bf j}}\approx 1$):

\vspace{1cm}
\begin{eqnarray}
&\Gamma_{\bf j}(T)&=\frac{1}{\tau_{{\bf j}}}=
\frac{2\pi}{\hbar}\sum\limits_{{\bf k},{\bf q}}|V({\bf q})|^{2}\delta\left(\varepsilon_{{\bf k}+{\bf
q}}+\varepsilon_{{\bf j}-{\bf q}}-\varepsilon_{{\bf k}}-\varepsilon_{{\bf j}}\right)\nonumber\\
&\times&\{(1-f_{{\bf j}})(1-f_{{\bf j-q}})(1-f_{{\bf k+q}})f_{{\bf k}}+f_{{\bf j}}f_{{\bf j-q}}f_{{\bf k+q}}(1-f_{{\bf k}})\}.\nonumber\\
\label{final}
\end{eqnarray}
This expression is one of our main results. In principle it allows for a direct calculation 
of lifetimes for any fermionic system with given quasi-particle dispersion relations and screened scattering term.
Very similar formulas can be found in textbooks in the context of
transport and relaxation, see, e.g.,  \cite{Jaeckle,Pines}. They are often
derived on the basis of an ad hoc application of Fermi's golden rule. A closer look reveals that such an expression can also be obtained from an 
analysis along the lines described in the introduction (''RPA-GW") by using the static-limit form of the screened interaction. Since our further 
quantitative determination only consists in a numerical evaluation of (\ref{final}) the outcome is equivalent to the one obtained by the above treatment. And since, as outlined in the following, (\ref{final}) is in accord with a standard scattering interpretation, obviously both, the projective and the above version of the many-body approach amount more or less to the counting of scattering events.\\
The contributions to the decay rate corresponding to $F_1$ and $F_2$
allow for an intuitive interpretation, at least for low temperatures.\\ 
$F_1$: This term accounts for the decay of an electron from a momentum
mode ${\bf j}$ above the Fermi sea, cf. Fig. (\ref{Collision}a). Due to the
factor $(1-f_{{\bf j}})$ it only significantly contributes to the
occupation number dynamics of such modes that are unoccupied in
equilibrium. Those occupation numbers may only deviate from
equilibrium towards an excess of electrons. According to the other three factors only those summands
contribute that correspond to the electron at ${\bf j}$ colliding with
an electron from within the Fermi sea ${\bf k}$, such that the
post-collision momenta ${\bf k+q}$, ${\bf j-q}$ lay in the unoccupied
region above the Fermi sea.\\
$F_2$: This term accounts for the decay of a hole from a momentum
mode ${\bf j}$ within  the Fermi sea, cf. Fig. (\ref{Collision}b). Due to the
factor $f_{{\bf j}}$ it only significantly contributes to the
occupation number dynamics of such modes that are occupied in
equilibrium. Those occupation numbers may only deviate from
equilibrium towards a shortage of electrons, i.e., holes. According to the other three factors only those summands
contribute that correspond to two electrons from within the Fermi sea
${\bf k+q}$, ${\bf j-q}$ colliding  such that one
post-collision momentum ${\bf k}$ lays in the unoccupied
region above the Fermi sea and the other lays exactly at ${\bf j}$ such as to
fill up the hole.\\

\section{Application to a  jellium model with screened interaction }   
\label{sec-2}
\begin{figure}
\centering
\includegraphics[width=80mm]{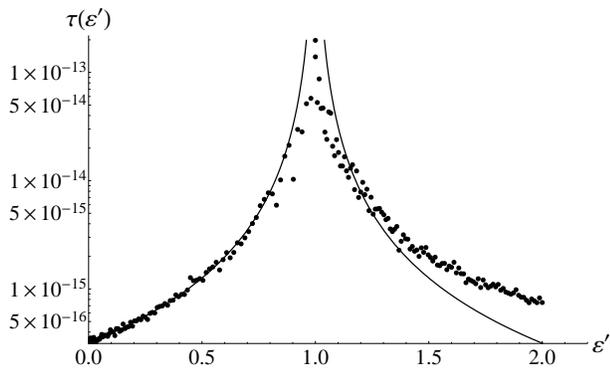}
\vspace*{1mm}
\caption{\label{Comparison2}Comparison of the logarithmic lifetimes of excited electrons (above $\varepsilon'= 1$)
and holes (below $\varepsilon'= 1$) in aluminium as arising from electron-electron scattering only. Displayed is a wide regime around the Fermi 
edge. Data are plotted over rescaled energy, $\varepsilon'=\varepsilon/\varepsilon_F$.
Displayed are results obtained
from Fermi-liquid theory as cited in (\cite{Quinn1}) (solid line, T=0K) and from numerical integration of (\ref{FinalRate})
(dots, T=10K). The number of
sample points for the Monte-Carlo integration  of (\ref{FinalRate}) is N$=10^{7}$ and $\sigma =1/10$.}
\end{figure}

In this section we now apply our result for the decay rates to a jellium modell featuring a Thomas-Fermi screened interaction. The latter will eventually be tuned to correspond to  aluminium. However, to repeat, the main intention of this work is not to calculate decay rates in aluminium with extreme precision, but to concretely demonstrate the feasibility of our method. The Hamiltonian 
of the model is given by:
\begin{eqnarray}
\hat{H}_{J}&=&\frac{\hbar^{2}}{2m_{e}}\sum\limits_{{\bf k}}k^{2}a_{{\bf k}}^{\dagger}a_{{\bf k}}\nonumber\\
&+&\frac{1}{2}\sum\limits_{{\bf 
k,l,q}}\frac{e^2}{\Omega\varepsilon_{0}({\bf q}^{2}+{\bf q}^{2}_{TF})}a_{{\bf
k+q}}^{\dagger}a_{{\bf l-q}}^{\dagger}a_{{\bf l}}a_{{\bf k}},\nonumber\\
\label{HamJel}
\end{eqnarray}
where $\Omega$ denotes the volume of the solid and ${\bf q}_{TF}$ the so called Thomas-Fermi wavenumber which is related to the Fermi wavevector and 
the Wigner-Seitz radius by $(q_{TF}/k_{F})^{2}=0.665r_{S}$ (with $r_{S}=(\frac{3}{4\pi\rho_{0}})^{\frac{1}{3}}\frac{1}{a_{0}}$, $a_{0}$ being the 
Bohr radius, $\varepsilon_{F}=(9\pi/4)^{\frac{2}{3}}\frac{1}{r_{S}^{2}}$[ryd], 
$k_{F}=(9\pi/4)^{\frac{1}{3}}\frac{1}{a_{0}r_{S}}$, $q_{TF}=(12/\pi)^{\frac{1}{3}}\frac{1}{a_{0}\sqrt{r_{S}}}$). Note that our model only comprises electrons, no phonons. Thus the result on the decay rate has to be compared to that part of the total decay rate that stems from electron-electron scattering only. In real aluminium there is evidence that the total lifetime is also significantly 
shortened due to electron-phonon scattering \cite{Bauer1,Nechaev}.
We apply now (\ref{final}) to this model which yields:
\begin{eqnarray}
\Gamma_{\bf j}(T)&=&\frac{1}{\tau_{{\bf j}}}=\frac{4\pi}{\hbar}\sum\limits_{{\bf k},{\bf q}}\left(\frac{e^{2}}{\Omega\varepsilon_{0}({\bf 
q}^{2}+{\bf 
q}_{TF}^{2})}\right)^{2}F({\bf k},{\bf 
q},{\bf j},T)\nonumber\\
&\times&\delta\left(\varepsilon_{{\bf k}+{\bf
q}}+\varepsilon_{{\bf j}-{\bf q}}-\varepsilon_{{\bf k}}-\varepsilon_{{\bf j}}\right)
\label{lebensdauer1}
\end{eqnarray}
where the auxiliary factor 2 arises from the fact that for each ${\bf k}$ we have two one-electron states (one for each spin)
which is taken into account by this additional factor. If we now replace the sums in 
(\ref{lebensdauer1}) by integrals by the rule $\sum\limits_{{\bf k}}f({\bf k})\rightarrow (2\pi)^{-3}\Omega\int d{\bf k}f({\bf k})$ we obtain for 
$\Gamma_{\bf j}(T)$:
\begin{eqnarray}
\Gamma_{\bf j}(T)
&=&\frac{4\Omega^{2}}{2^6\hbar\pi^{5}}\iint d{\bf k}d{\bf q}\left(\frac{e^{2}}{\Omega\varepsilon_{0}({\bf q}^{2}+{\bf
q}_{TF}^{2})}\right)^{2}F({\bf k},{\bf
q},{\bf j},T)\nonumber\\
\nonumber\\
&\times&\delta\left(\varepsilon_{{\bf k}+{\bf
q}}+\varepsilon_{{\bf j}-{\bf q}}-\varepsilon_{{\bf k}}-\varepsilon_{{\bf j}}\right)\nonumber\\
&=&\frac{e^{4}}{16\hbar\varepsilon_{0}^{2}\pi^{5}}\iint d{\bf k}d{\bf q}\frac{F({\bf k},{\bf
q},{\bf j},T)}{({\bf q}^{2}+{\bf
q}_{TF}^{2})^{2}}\nonumber\\
\nonumber\\
&\times&\delta\left(\varepsilon_{{\bf k}+{\bf
q}}+\varepsilon_{{\bf j}-{\bf q}}-\varepsilon_{{\bf k}}-\varepsilon_{{\bf j}}\right).\nonumber\\
\label{Rate2}
\end{eqnarray}
For the numerical calculation it is advantageous to transform the momenta to dimensionless parameter thus we introduce coordinates relative 
to the Fermi momentum: ${\bf k}\rightarrow k_{F}\cdot{\bf k'}$,
thus ${\bf q}\rightarrow k_{F}\cdot{\bf q'}$ and thus ${\bf j}\rightarrow k_{F}\cdot{\bf j'}$.
\begin{figure}
\centering
\includegraphics[width=80mm]{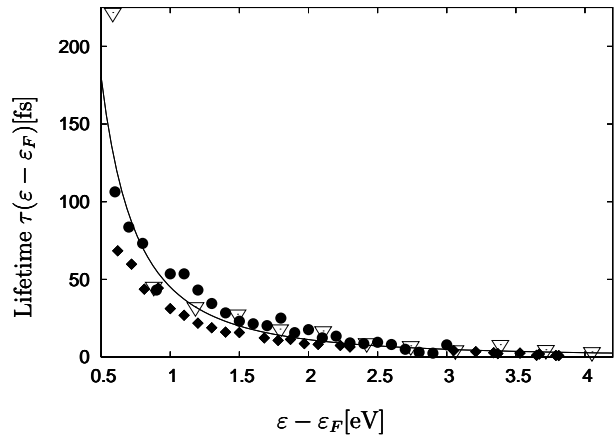}
\vspace*{1mm}
\caption{\label{Comparison} Comparison of the lifetimes of excited electrons in aluminium as arising from electron-electron scattering only. Displayed is the regime close to the Fermi edge. The open triangles correspond to data obtained from from the approach at hand, i.e., integration of  (\ref{FinalRate}) (T$=10$K).
The solid circles are experimental data corrected for transport effects taken from (\cite{Bauer1}) while the solid line is the analytical result 
from 
\cite{Quinn1}. 
Solid diamonds are the results of an DFT-GW calculation as shown in \cite{Echenique2}.
The number of sample points for the Monte-Carlo integration of (\ref{FinalRate}) is N=$2\cdot 10^{7}$, the broadening parameter is $\sigma$=1/25.}
\end{figure}
Applying all these substitutions to (\ref{Rate2}) leads to:
\begin{eqnarray}
\Gamma_{{\bf j}'}(T)&=&
\frac{m_{e}e^{4}}{8\hbar^{3}\varepsilon_{0}^{2}\pi^{5}}\iint d{\bf k}'d{\bf q}'\frac{F({\bf k}',{\bf
q}',{\bf j}',T)}{({\bf q}'^{2}+0.665r_{S})^{2}}\nonumber\\
\nonumber\\
&\times&\delta(({\bf k}'+{\bf q}')^{2}+({\bf j}'-{\bf q}')^{2}-{\bf k}'^{2}-{\bf j}'^{2}),
\end{eqnarray}
with $f_{{\bf k}'}=(\exp\left(\beta\varepsilon_{F}({\bf k}'^{2}-1)\right)+1)^{-1}$ and $m_{e}=9.1\cdot 10^{-31}$kg the free electron mass.\\
For the numerical evaluation of the last expression we approximate the
delta distribution by a suitable non-singular, e.g., Gaussian-type function:
\begin{eqnarray}
\delta_{\sigma}(\omega)\approx \frac{1}{\sigma\sqrt{2\pi}}e^{-\frac{\omega^{2}}{2\sigma^{2}}},
\end{eqnarray}
where $\sigma$ denotes the standard deviation which has to be in some
sense small. More details on this somewhat subtle approximation are
given below. Thus we get the following integral for the rate:
\begin{eqnarray}
\Gamma_{{\bf j}'}(T)
&=&\frac{1.05\text{fs}^{-1}}{\sigma}\cdot\iint d{\bf k}'d{\bf q}'\frac{F({\bf k}',{\bf
q}',{\bf j}',T)}{({\bf q}'^{2}+0.665r_{S})^{2}}\nonumber\\
&\times&\exp \left(\frac{1}{2\sigma^{2}}\left(({\bf k}'+{\bf q}')^{2}+({\bf j}'-{\bf q}')^{2}-{\bf 
k}'^{2}-{\bf 
j}'^{2}\right)^{2}\right).\nonumber\\
\label{FinalRate}
\end{eqnarray}
For aluminium we choose $r_S=2.07$.
The six dimensional integrals are solved numerically without any
further simplification using a standard Monte-Carlo package as implemented in 
the \emph{Mathematica} code. Of course this specific integral could be evaluated 
in other ways, however, to demonstrate the feasibility of our approach in general we proceed as indicated.\\ 
As one can see in Fig. \ref{Comparison} there is rather good
agreement between our results, other theoretical approaches and experiment. 
Our data is denoted by open triangles. The solid line corresponds to the many-body approach 
based on jellium \cite{Quinn1} as outlined in the introduction. The solid diamonds denote 
the result of a more sophisticated many-body approach which takes the lattice into 
account and exploits density functional theory \cite{Echenique2}. The solid circles indicate the  
parts of the measured decay rates that are attributed to direct electron-electron scattering, i. e., 
after removal of transport effects according to \cite{Bauer1}. Fig.\ref{Comparison2} shows the 
analytic result from \cite{Quinn1} ($263r_{S}^{-5/2}(\varepsilon-\varepsilon_{F})^{-2}[\text{eV}]^{2}[\text{fs}]$), 
which is supposed to be valid close to the 
Fermi edge, boldly continued to all energies (solid line). Furthermore results of our approach 
for all energies are displayed (dots). Obviously there are deviations for electrons at higher 
energies while the agreement remains very good in the limit of ``low-energy holes'' .\\ 
However, a comment should be added here. For this more or
less realistic model we get lifetimes on the order of some
femtoseconds. The decaytime of the correlation function (\ref{Rate}) is,
very roughly, on the order of $h/\epsilon_{max}$, with $\epsilon_{max}$
being the bandwith. For about 10eV this yields ca. half a 
femtosecond. Thus the separation of those timescales, which has been
mentioned in Sec. \ref{sec-2} as a criterion for the truncation
performed above, is not as clear as often in other fields, such as,
e.g., quantum optics. This indicates that such models, at short lifetimes, are barely 
in the Markovian, weak coupling regime and hence memory effects and/or
higher orders may have significant influence.\\
To the choice of $\sigma$: Obviously a smaller $\sigma$ leads to a
better approximation of the $\delta$-function which should be the
correct weight distribution at least in the long time limit. However,
recall the above discussion of the
time-independence of the decay rate. For analogous reasons larger $\sigma$ should leave the
result unaltered, as long as $\sigma$ remains small enough to allow
for a linearization of the dispersion relations on the scale of
$\sigma$. A large $\sigma$ is numerically favorable since the larger
$\sigma$ is, the larger will be the fraction of the Monte Carlo points
that significantly contribute to the integral. And of course this
yields a decreasing statistical error. Thus, for a given statistical
integration error, a larger  $\sigma$ simply implys a longer computing
time. Hence finding the best $\sigma$ is an optimization process that
should be done carefully. However, to name a number, the computation time for one of the lifetimes as displayed in Figs. \ref{Comparison}, \ref{Comparison2} is about an hour.

\section{Summary, conclusion and outlook}   
\label{sec-3}

In this paper we considered the lifetimes of (quasi-)particles or
holes in interacting quantum gases (only electronic part), using a projection operator
technique. This yields a formula for the decay rates into  which
essentially the pertinent, effective quasi-particle dispersion relations of the particles and their screened 
interactions enter. This formula turns out to be in accord with an expression that may be found from 
a certain implementation of the self-energy formalism . The rates are eventually given 
in terms of  integrals which can be cast into a form which is well suited for a Monte Carlo integration 
scheme. While this work essentially aims at demonstrating the feasibility of this approach in general, 
the method has been concretely applied to a jellium model featuring a Thomas-Fermi screened interaction 
(tuned for aluminium) as a simple example. Here it yields reasonable results while requiring moderate 
computational effort. This motivates an application of the approach to more complex systems. However, the results on life and correlation 
times indicate that such systems are, for short lifetimes (high electron energies, etc), barely Markovian and thus the
decay may not even be strictly exponential. This hints at a necessity to include higher order terms in future investigations in this regime.\\
The approach at hand aimed at generating an autonomous, linear equation of motion for a single
electron occupation number (\ref{rateequa}). However a slight modification of the projection used here may 
directly yield linear equation of motion for all electron occupation numbers, i.e.,  a linearized 
Boltzmann equation. As wellknown, the latter is a traditional starting point to investigate, e.g., 
transport properties. To those ends one would use a projection very much like the one discussed 
here (\ref{proj}) but summed over all occupation numbers $j$. The reasonable results on lifetimes presented 
in this work may be viewed to encourage further investigations in that direction.

\begin{appendix}
\label{sec-3}
\section{}
In this section we show the derivation of (\ref{final}).
The main work is to exploit the two commutators and finally the trace. First we exploit the commutator $[\hat{V}(0),\Delta_{{\bf j}}]$:
\begin{eqnarray*}
[\hat{V}(0),\Delta_{{\bf j}}]&=&\frac{1}{2}\sum\limits_{{\bf
k},{\bf l},{\bf q}}V({\bf q})[a_{{\bf k+q}}^{\dagger}a_{{\bf l-q}}^{\dagger}a_{{\bf l}}a_{{\bf k}},a_{\bf j}^{\dagger}a_{\bf j}-f_{{\bf 
j}}]\nonumber\\
&=&\frac{1}{2}\sum\limits_{{\bf
k},{\bf l},{\bf q}}V({\bf q})[a_{{\bf k+q}}^{\dagger}a_{{\bf l-q}}^{\dagger}a_{{\bf l}}a_{{\bf k}},a_{\bf j}^{\dagger}a_{\bf j}].\nonumber
\end{eqnarray*}
Since the commutator is zero for ${\bf j}\neq{\bf k+q},{\bf l-q},{\bf k},{\bf l}$ we just have to regard cases where one of the indices is equal to 
{\bf j} and note that $a_{{\bf i}}a_{{\bf i}}^{\dagger}a_{{\bf i}}=a_{{\bf i}}$, $a_{{\bf i}}^{\dagger}a_{{\bf 
i}}^{\dagger}a_{{\bf 
i}}=0$. From this follows that:
\begin{eqnarray}
[\hat{V}(0),\Delta_{{\bf j}}]&=&\frac{1}{2}\sum\limits_{{\bf k},{\bf q}}V({\bf q})a_{{\bf k+q}}^{\dagger}a_{{\bf j-q}}^{\dagger}a_{{\bf 
k}}a_{{\bf j}}\nonumber\\
&-&\frac{1}{2}\sum\limits_{{\bf l},{\bf q}}V({\bf q})a_{{\bf j+q}}^{\dagger}a_{{\bf l-q}}^{\dagger}a_{{\bf
l}}a_{{\bf j}}\nonumber\\
&-&\frac{1}{2}\sum\limits_{{\bf k},{\bf q}}V({\bf q})a_{{\bf j}}^{\dagger}a_{{\bf k}}^{\dagger}a_{{\bf
j-q}}a_{{\bf k+q}}\nonumber\\
&+&\frac{1}{2}\sum\limits_{{\bf l},{\bf q}}V({\bf q})a_{{\bf j}}^{\dagger}a_{{\bf l}}^{\dagger}a_{{\bf
l-q}}a_{{\bf j+q}}.
\end{eqnarray}
With suitable index shifts and the fermionic commutator relations we finally obtain for the commutator
\begin{eqnarray}
[\hat{V}(0),\Delta_{{\bf j}}]=\sum\limits_{{\bf k},{\bf q}}V({\bf q})\left(a_{{\bf k+q}}^{\dagger}a_{{\bf j-q}}^{\dagger}a_{{\bf
k}}a_{{\bf j}}-a_{{\bf j}}^{\dagger}a_{{\bf k}}^{\dagger}a_{{\bf j-q}}a_{{\bf k+q}}\right).\nonumber\\
\label{com1}
\end{eqnarray}
Now we deal with the second commutator $[\hat{V}(\tau),\tilde{\varrho}\otimes\Delta_{{\bf
j}}]$ where we first regard the case ${\bf j}\neq{\bf k+q},{\bf l-q},{\bf k},{\bf l}$ (we abbreviate $g_{{\bf i}}:=1-f_{{\bf i}}$):
\begin{eqnarray}
&~& [\hat{V}(\tau),\tilde{\varrho}\otimes\Delta_{{\bf j}}]^{\neq{\bf j}}\nonumber\\
&=&\frac{1}{2}\sum\limits_{{\bf
k},{\bf l},{\bf q}}V({\bf q})[a_{{\bf k+q}}^{\dagger}(\tau)a_{{\bf l-q}}^{\dagger}(\tau)a_{{\bf l}}(\tau)a_{{\bf 
k}}(\tau),\tilde{\varrho}_{s}\otimes\Delta_{{\bf
j}}]\nonumber\\
&=&\frac{1}{2}\sum\limits_{{\bf
k},{\bf l},{\bf q}}V({\bf q})e^{\frac{\imath}{\hbar}\left(\varepsilon_{{\bf k}+{\bf q}}+\varepsilon_{{\bf l}-{\bf
q}}-\varepsilon_{{\bf k}}-\varepsilon_{{\bf l}}\right)\tau}\nonumber\\
&\times&[a_{{\bf 
k+q}}^{\dagger}a_{{\bf l-q}}^{\dagger}a_{{\bf l}}a_{{\bf
k}},\bigotimes\limits_{{\bf i}\neq {\bf j}}\left(f_{{\bf i}}a_{\bf i}^{\dagger}a_{\bf i}+g_{{\bf i}}a_{\bf i}a_{\bf 
i}^{\dagger}\right)\otimes\Delta_{{\bf j}}],
\nonumber
\end{eqnarray}
where the annihilation and creation operators act only on the respective subspaces of the tensorproduct of the single density operators.
With $a_{{\bf i}}a_{{\bf i}}a_{{\bf i}}^{\dagger}=0$, $a_{{\bf i}}^{\dagger}a_{{\bf i}}a_{{\bf i}}^{\dagger}=a_{{\bf i}}^{\dagger}$, the rules above
and $u(\tau):=\exp\left(\frac{\imath}{\hbar}\left(\varepsilon_{{\bf k}+{\bf q}}+\varepsilon_{{\bf l}-{\bf q}}-\varepsilon_{{\bf 
k}}-\varepsilon_{{\bf l}}\right)\tau\right)$ it follows:
\begin{eqnarray}
&~&[\hat{V}(\tau),\tilde{\varrho}\otimes\Delta_{{\bf j}}]^{\neq{\bf j}}\nonumber\\
&=&\frac{1}{2}\sum\limits_{{\bf
k},{\bf l},{\bf q}}V({\bf q})u(\tau)\left(f_{{\bf k}}f_{{\bf l}}g_{{\bf l-q}}g_{{\bf k+q}}-g_{{\bf k}}g_{{\bf l}}f_{{\bf l-q}}f_{{\bf 
k+q}}\right)\nonumber\\
&\times&a_{{\bf k+q}}^{\dagger}a_{{\bf l-q}}^{\dagger}a_{{\bf l}}a_{{\bf k}}\bigotimes\limits_{\substack{{\bf i}\neq {\bf j},{\bf k},\\{\bf l},{\bf 
k+q},\\{\bf 
l-q}}}
\varrho_{{\bf i}}^{eq}\otimes\Delta_{{\bf j}}.\nonumber\\
\label{neqj1}
\end{eqnarray}
For the cases where one of the indices ${\bf k+q}$, ${\bf l-q}$, ${\bf k}$, ${\bf l}$ is equal ${\bf j}$ we obtain analogous:
\begin{widetext}
\begin{eqnarray}
[\hat{V}(\tau),\tilde{\varrho}\otimes\Delta_{{\bf j}}]^{={\bf j}}&=&
\frac{1}{2}\sum\limits_{{\bf k},{\bf q}}V({\bf q})a_{{\bf k+q}}^{\dagger}(\tau)a_{{\bf j-q}}^{\dagger}(\tau)a_{{\bf k}}(\tau)a_{{\bf
j}}(\tau)\bigotimes\limits_{{\substack{{\bf i}\neq {\bf j},{\bf k},\\{\bf j-q},\\{{\bf k+q}}}}}\varrho_{{\bf
i}}^{eq}
\left(g_{{\bf j}}g_{{\bf j-q}}g_{{\bf k+q}}f_{{\bf k}}+f_{{\bf j}}f_{{\bf j-q}}f_{{\bf k+q}}g_{{\bf k}}\right)\nonumber\\
&+&\frac{1}{2}\sum\limits_{{\bf l},{\bf q}}V({\bf q})a_{{\bf j+q}}^{\dagger}(\tau)a_{{\bf l-q}}^{\dagger}(\tau)a_{{\bf j}}(\tau)a_{{\bf
l}}(\tau)\bigotimes\limits_{{\substack{{\bf i}\neq {\bf j},{\bf l},\\{\bf l-q},\\{\bf j+q}}}}\varrho_{{\bf
i}}^{eq}
\left(g_{{\bf j}}g_{{\bf l-q}}g_{{\bf j+q}}f_{{\bf l}}+f_{{\bf j}}f_{{\bf j+q}}f_{{\bf l-q}}g_{{\bf l}}\right)\nonumber\\
&-&\frac{1}{2}\sum\limits_{{\bf k},{\bf q}}V({\bf q})a_{{\bf k+q}}^{\dagger}(\tau)a_{{\bf j}}^{\dagger}(\tau)a_{{\bf k}}(\tau)a_{{\bf
j+q}}(\tau)\bigotimes\limits_{{\substack{{\bf i}\neq {\bf j},{\bf j+q},\\{\bf k},{\bf k+q}}}}\varrho_{{\bf i}}^{eq}
\left(g_{{\bf j+q}}g_{{\bf k}}g_{{\bf j}}f_{{\bf k+q}}+f_{{\bf j}}f_{{\bf k}}f_{{\bf j+q}}g_{{\bf k+q}}\right)\nonumber\\
&-&\frac{1}{2}\sum\limits_{{\bf l},{\bf q}}V({\bf q})a_{{\bf j}}^{\dagger}(\tau)a_{{\bf l-q}}^{\dagger}(\tau)a_{{\bf j-q}}(\tau)a_{{\bf
l}}(\tau)\bigotimes\limits_{{\substack{{\bf i}\neq {\bf j},{\bf j-q},\\{\bf l},{\bf l-q}}}}\varrho_{{\bf i}}^{eq}
\left(g_{{\bf j-q}}g_{{\bf l}}g_{{\bf j}}f_{{\bf l-q}}+f_{{\bf j}}f_{{\bf l}}f_{{\bf j-q}}g_{{\bf l-q}}\right).\nonumber\\
\end{eqnarray}
\end{widetext}
Again with suitable indexshifts and substitutions it follows for the commutator:
\begin{widetext}
\begin{eqnarray}
[\hat{V}(\tau),\tilde{\varrho}\otimes\Delta_{{\bf j}}]^{={\bf j}}&=&
\sum\limits_{{\bf k},{\bf q}}V({\bf q})\left(a_{{\bf k+q}}^{\dagger}(\tau)a_{{\bf j-q}}^{\dagger}(\tau)a_{{\bf k}}(\tau)a_{{\bf
j}}(\tau)-a_{{\bf j}}^{\dagger}(\tau)a_{{\bf k}}^{\dagger}(\tau)a_{{\bf j-q}}(\tau)a_{{\bf k+q}}(\tau)\right)\nonumber\\
&\times&\left(g_{{\bf j}}g_{{\bf j-q}}g_{{\bf k+q}}f_{{\bf k}}+f_{{\bf j}}f_{{\bf j-q}}f_{{\bf k+q}}g_{{\bf k}}\right)
\bigotimes\limits_{{\substack{{\bf i}\neq {\bf j},{\bf k},\\{\bf j-q},\\{{\bf k+q}}}}}\varrho_{{\bf i}}^{eq}
\end{eqnarray}
\end{widetext}
\begin{widetext}
Now we exploit the trace:
\begin{eqnarray}
C(\tau)&=&\text{Tr}\left\{[\hat{V}(\tau),\tilde{\varrho}\otimes\Delta_{{\bf j}}][\hat{V}(0),\Delta_{{\bf j}}]\right\}
=\text{Tr}\left\{\left([\hat{V}(\tau),\tilde{\varrho}\otimes\Delta_{{\bf j}}]^{\neq{\bf j}}+
[\hat{V}(\tau),\tilde{\varrho}\otimes\Delta_{{\bf j}}]^{={\bf j}}\right)[\hat{V}(0),\Delta_{{\bf 
j}}]\right\}\nonumber\\
&=& \underbrace{\text{Tr}\left\{[\hat{V}(\tau),\tilde{\varrho}\otimes\Delta_{{\bf j}}]^{\neq{\bf j}}[\hat{V}(0),\Delta_{{\bf
j}}]\right\}}_{A(\tau)}
+\underbrace{\text{Tr}\left\{[\hat{V}(\tau),\tilde{\varrho}\otimes\Delta_{{\bf j}}]^{={\bf j}}[\hat{V}(0),\Delta_{{\bf
j}}]\right\}}_{B(\tau)}\nonumber,
\end{eqnarray}
where we split it into two parts and exploit them respectively.
%
\begin{eqnarray}
A(\tau)&=&\frac{1}{2}\sum\limits_{\substack{{\bf
k},{\bf l},{\bf q}\\{\bf x},{\bf y}}}V({\bf q})V({\bf y})u(\tau)G({\bf k},{\bf l},{\bf q},T)\text{Tr}\left\{a_{{\bf k+q}}^{\dagger}a_{{\bf 
l-q}}^{\dagger}a_{{\bf 
l}}a_{{\bf k}}\bigotimes\limits_{\substack{{\bf i}\neq {\bf j},{\bf k},\\{\bf l},{\bf
k+q},\\{\bf
l-q}}}\varrho_{{\bf i}}^{eq}\otimes\Delta_{{\bf j}}
\left(a_{{\bf x+y}}^{\dagger}a_{{\bf j-y}}^{\dagger}a_{{\bf
x}}a_{{\bf j}}-a_{{\bf j}}^{\dagger}a_{{\bf x}}^{\dagger}a_{{\bf j-y}}a_{{\bf x+y}}\right)\right\}\nonumber\\
&=&\frac{1}{2}\sum\limits_{\substack{{\bf
k},{\bf l},{\bf q}\\{\bf x},{\bf y}}}V({\bf q})V({\bf y})u(\tau)G({\bf k},{\bf l},{\bf q},T)\underbrace{\text{Tr}\left\{a_{{\bf 
k+q}}^{\dagger}a_{{\bf
l-q}}^{\dagger}a_{{\bf
l}}a_{{\bf k}}\bigotimes\limits_{\substack{{\bf i}\neq {\bf j},{\bf k},\\{\bf l},{\bf
k+q},\\{\bf
l-q}}}\varrho_{{\bf i}}^{eq}\otimes\Delta_{{\bf j}}
a_{{\bf x+y}}^{\dagger}a_{{\bf j-y}}^{\dagger}a_{{\bf
x}}a_{{\bf j}}\right\}}_{\text{I}}\nonumber\\
&-&
\frac{1}{2}\sum\limits_{\substack{{\bf
k},{\bf l},{\bf q}\\{\bf x},{\bf y}}}V({\bf q})V({\bf y})u(\tau)G({\bf k},{\bf l},{\bf q},T)\underbrace{\text{Tr}\left\{a_{{\bf 
k+q}}^{\dagger}a_{{\bf
l-q}}^{\dagger}a_{{\bf
l}}a_{{\bf k}}\bigotimes\limits_{\substack{{\bf i}\neq {\bf j},{\bf k},\\{\bf l},{\bf
k+q},\\{\bf
l-q}}}\varrho_{{\bf i}}^{eq}\otimes\Delta_{{\bf j}}
a_{{\bf j}}^{\dagger}a_{{\bf x}}^{\dagger}a_{{\bf j-y}}a_{{\bf x+y}}\right\}}_{\text{II}}
\label{atau}
\end{eqnarray}
with $G({\bf k},{\bf l},{\bf q},T)=f_{{\bf k}}f_{{\bf l}}g_{{\bf l-q}}g_{{\bf k+q}}-g_{{\bf k}}g_{{\bf l}}f_{{\bf l-q}}f_{{\bf
k+q}}$. We focus now on the traces. For taking the trace we use the occupation number representation.\\
\underline{{\bf I}}:
\begin{eqnarray}
\text{Tr}\left\{a_{{\bf
k+q}}^{\dagger}a_{{\bf
l-q}}^{\dagger}a_{{\bf
l}}a_{{\bf k}}\underbrace{\bigotimes\limits_{\substack{{\bf i}\neq {\bf j},{\bf k},\\{\bf l},{\bf
k+q},\\{\bf
l-q}}}\varrho_{{\bf i}}^{eq}}_{\Xi}\otimes\Delta_{{\bf j}}
a_{{\bf x+y}}^{\dagger}a_{{\bf j-y}}^{\dagger}a_{{\bf
x}}a_{{\bf j}}\right\}
=
\sum\limits_{n_{1},\dots,n_{r}}\langle n_{1},\dots,n_{r}|
a_{{\bf
k+q}}^{\dagger}a_{{\bf
l-q}}^{\dagger}a_{{\bf
l}}a_{{\bf k}}\cdot\Xi\otimes\Delta_{{\bf j}}\cdot
a_{{\bf x+y}}^{\dagger}a_{{\bf j-y}}^{\dagger}a_{{\bf
x}}a_{{\bf j}}
|n_{1},\dots,n_{r}\rangle\nonumber\\
\end{eqnarray}
Since under consideration no one of the indices ${\bf k}, {\bf l}, {\bf k+q}, {\bf 
l-q}$ is equal ${\bf j}$ this trace is zero for cases with ${\bf y}\neq0$ which is valid for II also.
The case ${\bf y}=0$ must be analyzed independent.\\
\underline{{\bf I}} for  ${\bf y}=0$ (now we write down the sum again) we have:
\begin{eqnarray}
\frac{1}{2}\sum\limits_{\substack{{\bf
k},{\bf l}\\{\bf q},{\bf x}}}V({\bf q})V({\bf 0})u(\tau)G({\bf k},{\bf l},{\bf q},T)\text{Tr}\left\{a_{{\bf
k+q}}^{\dagger}a_{{\bf
l-q}}^{\dagger}a_{{\bf
l}}a_{{\bf k}}\cdot\Xi\otimes\Delta_{{\bf j}}\cdot
a_{{\bf x}}^{\dagger}a_{{\bf j}}^{\dagger}a_{{\bf
x}}a_{{\bf j}}\right\}.
\end{eqnarray}
Here we have two summands: the case where ${\bf k+q}={\bf l}$ and ${\bf l+q}={\bf l}\rightarrow {\bf q}=0$:
\begin{eqnarray}
&~&\frac{1}{2}\sum\limits_{\substack{{\bf
k},{\bf l}\\{\bf q},{\bf x}}}V({\bf q})V({\bf 0})u(\tau)G({\bf k},{\bf l},{\bf q},T)\text{Tr}\left\{a_{{\bf
k+q}}^{\dagger}a_{{\bf
l-q}}^{\dagger}a_{{\bf
l}}a_{{\bf k}}\cdot\Xi\otimes\Delta_{{\bf j}}\cdot
a_{{\bf x}}^{\dagger}a_{{\bf j}}^{\dagger}a_{{\bf
x}}a_{{\bf j}}\right\}=\nonumber\\
&~&\frac{1}{2}\sum\limits_{\substack{{\bf
k},{\bf q},{\bf x}}}V({\bf q})V({\bf 0})u(\tau)G({\bf k},{\bf k+q},{\bf q},T)\text{Tr}\left\{a_{{\bf
k+q}}^{\dagger}a_{{\bf
k}}^{\dagger}a_{{\bf
k+q}}a_{{\bf k}}\cdot\Xi\otimes\Delta_{{\bf j}}\cdot
a_{{\bf x}}^{\dagger}a_{{\bf j}}^{\dagger}a_{{\bf
x}}a_{{\bf j}}\right\}+\nonumber\\
&~&\frac{1}{2}\sum\limits_{\substack{{\bf
k},{\bf l},{\bf x}}}V({\bf 0})V({\bf 0})u(\tau)G({\bf k},{\bf l},{\bf 0},T)\text{Tr}\left\{a_{{\bf
k}}^{\dagger}a_{{\bf
l}}^{\dagger}a_{{\bf
l}}a_{{\bf k}}\cdot\Xi\otimes\Delta_{{\bf j}}\cdot
a_{{\bf x}}^{\dagger}a_{{\bf j}}^{\dagger}a_{{\bf
x}}a_{{\bf j}}\right\}=0,\nonumber\\
\end{eqnarray} 
since $G({\bf k},{\bf k+q},{\bf q},T)=G({\bf k},{\bf l},{\bf 0},T)=0$. For II the argumentation is analogous.
Thus there is left just one more possibility: the case ${\bf j=x+y}$ for which we obtain from (\ref{atau}):
\begin{eqnarray}
&~&\frac{1}{2}\sum\limits_{\substack{{\bf
k},{\bf l},{\bf q},{\bf x}}}V({\bf q})V({\bf y})u(\tau)G({\bf k},{\bf l},{\bf q},T)\text{Tr}\left\{a_{{\bf
k+q}}^{\dagger}a_{{\bf
l-q}}^{\dagger}a_{{\bf
l}}a_{{\bf k}}\cdot\Xi\otimes\Delta_{{\bf j}}\cdot
a_{{\bf j}}^{\dagger}a_{{\bf j-y}}^{\dagger}a_{{\bf
j-y}}a_{{\bf j}}\right\}\nonumber\\
&-&
\frac{1}{2}\sum\limits_{\substack{{\bf
k},{\bf l},{\bf q},{\bf x}}}V({\bf q})V({\bf y})u(\tau)G({\bf k},{\bf l},{\bf q},T)\text{Tr}\left\{a_{{\bf
k+q}}^{\dagger}a_{{\bf
l-q}}^{\dagger}a_{{\bf
l}}a_{{\bf k}}\cdot\Xi\otimes\Delta_{{\bf j}}\cdot
a_{{\bf j}}^{\dagger}a_{{\bf j-y}}^{\dagger}a_{{\bf j-y}}a_{{\bf j}}\right\}=0,
\end{eqnarray}
so that finally follows that $A(\tau)=0$.
\vspace{1cm}
\end{widetext}
\begin{widetext}
For $B(\tau)$ we have:
\begin{eqnarray}
B(\tau)&=&\sum\limits_{\substack{{\bf
k},{\bf q}\\{\bf x},{\bf y}}}V({\bf q})V({\bf y})F({\bf k},{\bf q},{\bf j},T)\text{Tr}\left\{
\left(a_{{\bf k+q}}^{\dagger}(\tau)a_{{\bf j-q}}^{\dagger}(\tau)a_{{\bf k}}(\tau)a_{{\bf
j}}(\tau)-a_{{\bf j}}^{\dagger}(\tau)a_{{\bf k}}^{\dagger}(\tau)a_{{\bf j-q}}(\tau)a_{{\bf k+q}}(\tau)\right)\bigotimes\limits_{{\substack{{\bf 
i}\neq {\bf j},{\bf k},\\{\bf j-q},\\{{\bf k+q}}}}}\varrho_{{\bf i}}^{eq}
\right.\nonumber\\
&\times&\left.\left(a_{{\bf x+y}}^{\dagger}a_{{\bf j-y}}^{\dagger}a_{{\bf
x}}a_{{\bf j}}-a_{{\bf j}}^{\dagger}a_{{\bf x}}^{\dagger}a_{{\bf j-y}}a_{{\bf x+y}}\right)\right\}\nonumber\\
\nonumber\\
&=&-2\sum\limits_{\substack{{\bf
k},{\bf q}}}|V({\bf q})|^{2}\left(g_{{\bf j}}g_{{\bf j-q}}g_{{\bf k+q}}f_{{\bf k}}+f_{{\bf j}}f_{{\bf j-q}}f_{{\bf k+q}}g_{{\bf k}}\right)
\cos\left((\omega_{{\bf k}+{\bf q}}+\omega_{{\bf j}-{\bf q}}-\omega_{{\bf k}}-\omega_{{\bf
j}})\tau\right),
\end{eqnarray}
\end{widetext} 
from this follows (\ref{final}).
\end{appendix}
\begin{acknowledgments}
We thank K. B\"arwinkel and M. Rohlfing for
fruitful discussions. Financial support by the Deutsche Forschungsgemeinschaft and the Graduate College 695
``Nonlinearities of optical Materials'' is greatfully acknowledged.
\end{acknowledgments}

\bibliographystyle{IEEE}
\bibliography{myrefs}

\end{document}